\documentstyle[aps] {revtex}
\tighten
\twocolumn
\input{psfig.sty}
\begin{document}
\draft
\title{Dynamics of a single electron in the disordered Holstein model}
\author{F.X.~Bronold}
\address{Institut f\"ur Theoretische Physik, Otto-von-Guericke-Universit\"at Magdeburg\\
Universit\"atsplatz 2, PF 4120, 39016 Magdeburg, Germany}
\author{A.~Saxena and A.R.~Bishop}
\address{Theoretical Division and Center for Nonlinear Studies,
Los Alamos National Laboratory, \\
Los Alamos, New Mexico 87545}
%
%
\maketitle

\begin{abstract}
We study, at zero temperature, the dynamics of a single electron in a 
Holstein model augmented by site-diagonal, binary-alloy type
disorder. The average over the phonon 
vacuum and the alloy configurations is performed within 
a generalized dynamical coherent potential approximation. 
We present numerical results for a Bethe lattice with 
infinite coordination number. 
In particular, we investigate, in the 
intermediate electron-phonon coupling regime, the spectral
and diffusion properties in the vicinity of the
high-energy edge of the 
lowest polaronic subband. 
To characterize 
the diffusion properties,  
we define a spectrally resolved delocalization time, 
which is, for a given energy, the characteristic time scale
on which the electron leaves a given site. We find the 
delocalization times substantially enhanced for states with a 
large phonon content, i.e., in the absence (presence) of 
alloy-type disorder at the high-energy edge(s) of the polaronic subband
(mini-subbands). According to their delocalization times, we 
discriminate 
between ``fast'' {\it quasi-particle-like} and 
``sluggish'' {\it defect-like} 
polaron states and qualitatively
address the issue of  
trapping of an electronic carrier.
\end{abstract}

\pacs{PACS numbers: 71.38.+i, 72.10.Di, 71.35.Aa}

\section{Introduction}

In nature there is an abundance  
of ``polaronic'' compounds in which, due to 
strong electron-phonon coupling, electrons and phonons lose, 
in a certain temperature and density range, their individual 
identity and new entities emerge: polarons \cite{M90}.  
Examples of current interest are,
among others, the high-temperature
superconducting perovskites \cite{htc}, the ``colossal'' magneto-resistance
manganites \cite{manganites}, and the titanates \cite{titanites}. 
The details of the actual materials notwithstanding, it is commonly assumed 
that, at least qualitatively, many
aspects of these compounds can be understood on the  
basis of the Holstein model \cite{H59} supplemented by various terms, such as, e.g., 
Coulomb repulsion, spin-orbit coupling, and specific forms of 
material imperfections. 

In its basic form
the Holstein model comprises a single tight-binding band locally 
coupled to dispersionless
optical phonons. 
Despite this simplicity, the properties of the Holstein model are not 
yet fully understood.  
Even in the extreme dilute limit, that is the case of a single polaron 
in an otherwise empty band, many basic questions remain, 
despite impressive theoretical progress made in recent years
\cite{dRL83,S96,WF97,BTB99,L88,C97,dMR97,WF98}. 

In particular, the notoriously difficult issue of 
trapping \cite{L88,C97,dMR97,WF98} 
is far from being completely resolved.  
The trapping properties are only understood for the polaron
{\it groundstate} of the ordered Holstein model. In an important paper,  
L\"owen \cite{L88} has rigorously shown that the groundstate 
of the ordered Holstein model has to be 
itinerant, thereby implying that the groundstate can be only 
trapped by an {\it external} process, e.g., driven by material imperfections. 
Although it is not clear whether L\"owen's proof 
also applies to {\it excited} states of the polaron, it is
generally assumed that {\it all} states in the ordered Holstein 
model are itinerant. Hence, if trapping occurs, it has to 
be externally driven. Indeed,
translational invariance of the Holstein model seems to force all 
(eigen)states to be itinerant. On the other hand, the non-existence 
of translational invariance does not necessarily prohibit itinerant 
states. For a single electron in a random potential, for example,  
itinerant states exist, separated 
from localized states by a ``mobility edge'', despite the absence of 
translational invariance. Likewise, as noted above,
translational invariance 
per se might be no guarantee that all polaron states 
are itinerant. That is, the  
important question about the trapping properties of {\it excited}
polaron states is essentially unanswered. 

Evidently, the broad theoretical task is to construct a rigorous 
``mobility-edge-theory'' for {\it excited} polaron states, 
treating extrinsic and (possibly) intrinsic trapping processes 
on an equal footing. In view of the complexity of this problem, 
we are not yet able to develop such a rigorous theory. Instead, 
we present in this paper, as a first step, an approximate investigation 
of the spectral and diffusion properties of that part of the
spectrum, where we expect a rigorous theory to predict a mobility
edge. In particular, we investigate, in the intermediate 
electron-phonon coupling regime
and with an approximate technique, the states comprising the 
high-energy edge of the lowest polaronic subband. The dispersion
is extremely flat in this part of the spectrum indicating, perhaps, 
already the tendency of (intrinsic) trapping. Definitely, very 
small amounts of disorder suffice to drive these states into the
trapped regime.

Our investigation is based on the Holstein model  
augmented by site-diagonal, binary-alloy type disorder. 
Measuring energy in units
of $2J$, where $J$ denotes the electronic (nearest-neighbor) 
hopping integral, the Hamiltonian for the
``Holstein alloy'' reads in the single-electron sector 
\begin{eqnarray}
H&=&\epsilon_B \sum_i |i><i|
-{1\over 2} \sum_{<ij>} |i><j|
\nonumber\\
&+&\delta\sum_i x_i |i><i|
\nonumber\\
&+& \Omega \sum_i b_i^\dagger b_i
-g\sum_i (b_i + b_i^\dagger) |i><i|.
\label{model}
\end{eqnarray}
Here, $|i>$ stands for the Wannier state on site
$i$ and $b_i [b_i^\dagger]$ annihilates [creates] an optical phonon
on site $i$. The phonon frequency and the electron-phonon 
coupling constant (in units of $2J$) are denoted by 
$\Omega$ and $g$, respectively.
Depending on the occupancy of the site, the on-site energy 
(in units of $2J$) is either
$\epsilon_A$ or $\epsilon_B$. The scattering strength
$\delta=\epsilon_A-\epsilon_B$ can be taken as a measure of the 
disorder. The independent
random variables $x_i$ are drawn from a bi-modal probability distribution
$p(x_i)=c\delta(x_i)+(1-c)\delta(x_i-1)$,  
with $0\le c \le 1$ ($c$ denotes the concentration of the B sites).       

Spectral properties of the model, such as the density of states, the self-energy, 
and the spectral function, can be deduced from the electronic two-point 
function, which at temperature $T=0$ and in Wannier representation reads
\begin{eqnarray}
{\cal G}_{ij}(z)=
<i|\bar{G}(z)|j>,
\label{Gij}
\end{eqnarray}
with an {\it averaged} one-resolvent 
\begin{eqnarray}
\bar{G}(z)=
\langle\langle(0|G(z)|0)\rangle\rangle.
\label{Gbar1}
\end{eqnarray}
Here, $G(z)=1/(z-H)$ denotes the one-resolvent and
$\langle\langle...\rangle\rangle$ and $(0|...|0)$ stand, respectively, 
for averages over the quenched disorder and the phonon vacuum. Henceforth, the term
``averaged'' implies, if not otherwise specified,
averaged over both quenched disorder and the phonon vacuum. 

From the spectral properties alone it is, in general, not 
possible to decide whether the states involved are itinerant 
(extended) or trapped (localized). A more reliable method
to discriminate between the two kinds of states is to 
investigate the return probability $P$, i.e., the
probability to find (in the long time limit) the electron on the
same site at which it was initially injected. This criterion 
has been used, for example, to characterize electronic states 
in a random potential \cite{EC72}. At $T=0$ and in 
Wannier representation, 
the correlation function, whose long time limit is the 
return probability, reads
\begin{eqnarray}
P(t)=\langle\langle(0|<i|e^{iHt}|i><i|e^{-iHt}|i>|0)\rangle\rangle.
\end{eqnarray}  
Using Abel's theorem, $P$ can be conveniently expressed in terms 
of $p(2\eta)$, the Laplace transform of $1/2P(t/2)$ \cite{EC72}. 
Explicitly, 
\begin{eqnarray}
P&=&\lim_{t\rightarrow \infty} P(t)
\nonumber\\
&=&\lim_{\eta\rightarrow 0} 2\eta p(2\eta)
\nonumber\\
&=&\lim_{\eta\rightarrow 0} 2\eta \int_{-\infty}^{\infty}
{{d\omega}\over {2\pi}}
f(\omega-i\eta,\omega+i\eta),
\label{returnprob1}
\end{eqnarray}
where we have defined a four-point function
\begin{eqnarray}
f(z_1,z_2)=<i|\bar{K}(z_1,|i><i|,z_2)|i>,
\label{f1}
\end{eqnarray}
given in terms of an {\it averaged} two-resolvent, which (for 
an arbitrary electronic operator $O_e$) reads
\begin{eqnarray}
\bar{K}(z_1,O_e,z_2)=
\langle\langle(0|G(z_1)O_eG(z_2)|0)\rangle\rangle.
\label{kbar}
\end{eqnarray}

The behavior of the spectrally resolved return probability, 
\begin{eqnarray}
P(\omega,\eta)=2\eta f(\omega-i\eta,\omega+i\eta),
\label{returnprob2}
\end{eqnarray}    
in the limit $\eta\rightarrow 0$ allows one to 
distinguish trapped from itinerant states.
In particular, for a trapped 
polaronic defect state,
$\lim_{\eta\rightarrow 0}P(\omega,\eta)$ is finite,   
whereas for an itinerant polaronic quasi-particle state 
$\lim_{\eta\rightarrow 0}P(\omega,\eta)$ vanishes. 

The exact calculation of the averaged one- and two-resolvents for 
the Holstein alloy is a rather
formidable task, and, 
due to lack of rigorous mathematical methods, we are forced  
to adopt an effective single-site averaging procedure. Specifically,
we employ 
a generalized dynamical coherent potential approximation (DCPA)
\cite{S74,MH81,MH82,A88a,A88bA90,PLH84}, which is known 
to capture the essentials of the 
polaron formation process. 

Unfortunately, within the DCPA, trapped polaron states cannot be 
unambiguously identified, because  
$\lim_{\eta\rightarrow 0} P(\omega,\eta)$ vanishes for  
{\it all} energies $\omega$. Thus, 
rigorously speaking, the DCPA predicts all polaron states to be 
itinerant. Nevertheless, it is possible to give a qualitative 
discussion
of the trapping issue. In particular, an asymptotic 
analysis of $P(\omega,\eta)$ for {\it finite} $\eta$
allows to define a spectrally resolved delocalization time, 
which is the characteristic time scale on which, at a given
energy, the electron leaves a given site. According
to their delocalization times, it is then possibe to 
distinguish between ``fast'' {\it quasi-particle-like} 
and ``sluggish'', i.e., temporarily trapped,
{\it defect-like} polaron states. 

The organization of the rest of the paper is as follows. 
Section II contains  
a general description of the DCPA with details about
the calculation of 
the two- and the four-point functions defined above. 
Section III specializes the DCPA formalism to the 
Bethe lattice with infinite coordination number
and discusses representative numerical results for the spectral and 
diffusion properties,
respectively.  
Mathematical details not essential for the structure of the paper are 
relegated to 
two appendices. Finally, we conclude in section IV with a summary of 
key results and open questions.
 
\section{Dynamical Coherent Potential Approximation}

\subsection{Perspective}

In contrast to a classical alloy, where an electron  
experiences only elastic impurity scattering, the electron in a Holstein 
alloy is subject to elastic impurity and inelastic phonon scattering. 
Both scattering processes might possibly trap the electron.
It is therefore necessary to treat them simultaneously. In particular,
averages over the phonon vacuum and the alloy configurations need 
to be performed on an equal footing. 

A powerful tool to approximately perform these two averages is the 
effective single-site approximation, where one embeds an individual 
scatterer into an effective medium to be determined self-consistently
by enforcing the scattering from a single scatterer to vanish
on the average. The main advantage of this scheme is the non-perturbative 
treatment of on-site correlations. In particular, the 
non-perturbative treatment of on-site electron-phonon correlations
is an essential in order to capture polaron
formation. The main drawback, on the other hand, is 
the insufficient treatment of inter-site correlations. 
It is the latter which eventually makes 
$\lim_{\eta\rightarrow 0}P(\omega,\eta)$ to vanish for all
energies $\omega$, irrespective of the model parameters, and, 
thus, prevents a 
rigorous calculation of diffusion 
and trapping properties. However, as indicated in the Introduction,
an asymptotic analysis of $P(\omega,\eta)$ for finite $\eta$ 
partly compensates for this shortcoming.

In the case of elastic impurity scattering  
the effective single-site approximation yields the well-known 
coherent potential approximation (CPA)
\cite{S67,V68,V69}. The  
DCPA, originally developed by Sumi \cite{S74},
is a direct extension of the CPA to problems involving
{\it inelastic} scattering. It 
has been primarily used to investigate linear \cite{S74,MH81}
and nonlinear \cite{MH82,A88a,A88bA90} optical properties of 
polaron-excitons
in molecular crystals. However, it can be applied to 
any on-site inelastic 
scattering process. Paquet and Leroux-Hugon \cite{PLH84}
employed the DCPA, e.g., to investigate 
a single quantum particle subject to an on-site potential which fluctuates 
in time according to discrete or continuous Markov processes.

The CPA as well as the DCPA are mean field
approximations for a single electron subject to impurity or phonon 
scattering, respectively. The most general mean field 
approximation, applicable also to dense quantum systems with mutual 
inelastic scattering between its constituents, is the dynamical mean 
field approximation (DMFA) \cite{V93,P95,G96}. [Naturally, in the
limit of a single electron, interacting with a bath (of phonons 
or random scatterers), the DMFA reduces to the (D)CPA.]  
It can be shown, assuming that the  
nearest-neighbor hopping integral is scaled by a factor $1/\sqrt{Z}$,  
where $Z$ denotes the coordination number of the lattice, 
that the DMFA  
provides an {\it exact} solution of the original model in the limit 
of $Z \rightarrow \infty$. Accordingly, 
the DMFA [i.e., in the respective limits, the (D)CPA]
can be either considered as an exact 
theory for lattices with $Z=\infty$ or as an approximate theory for 
lattices with finite $Z$.  

The DMFA has been employed by Ciuchi et al. \cite{C97} to study, in 
various electron-phonon coupling regimes, the 
groundstate and the spectral properties of a single electron in the 
{\it ordered} 
Holstein model. As indicated above, for a single electron,
the DMFA reduces to the DCPA.
The investigations of Sumi and Ciuchi et al. are therefore closely related.

Postponing to section III a discussion of how our 
investigation complements the DCPA work by Sumi \cite{S74} and the 
DMFA work by Ciuchi et al. \cite{C97}, we shall now present
a derivation of the generalized DCPA formalism as needed to 
determine two-point 
and four-point functions for the Holstein alloy. 

\subsection{Calculation of $\bar{G}(z)$}

In the spirit of the CPA we introduce an effective medium described by an
effective one-resolvent
\begin{eqnarray}
G_{eff}(z)={1\over{z-H_{eff}(z)}},
\label{geff}
\end{eqnarray}
with 
\begin{eqnarray}
H_{eff}(z)&=&H_B+\Sigma(z),
\end{eqnarray}
and 
\begin{eqnarray}
H_B&=&\epsilon_B \sum_i |i><i|
-{1\over 2} \sum_{<ij>} |i><j|
\nonumber\\
&+& \Omega\sum_i b_i^\dagger b_i.
\end{eqnarray}

The unknown energy-dependent 
coherent potential (self-energy) operator $\Sigma(z)$ defined with respect to the 
reference Hamiltonian $H_B$ will be specified by enforcing a self-consistency 
condition, namely the averaged one-resolvent is forced to be equal to the 
averaged effective 
one-resolvent, i.e.,
\begin{eqnarray}
\bar{G}(z)=
(0|G_{eff}(z)|0),
\label{sce1}
\end{eqnarray}
where we have used the configuration independence of $G_{eff}(z)$.

In the above equations, the energy variable $z$
denotes the {\it total} energy of the coupled electron-phonon system and not 
just the electron energy.
The coherent potential seen by the electron,  
on the other hand, depends of course only on the electron
energy. Therefore, the energy argument of the coherent potential 
is the total energy $z$ minus the
lattice energy, which is, neglecting the zero point motion of the phonons, 
just the total number of phonons in the system times the phonon frequency. 
A convenient way to ensure that the coherent potential is only a function of 
the electron energy is to define,  
following ref. \cite{PLH84}, a set of projection operators $P_q$, 
which 
project onto the phonon subspace with a total number of $q$ phonons, and
to write for the coherent potential operator
\begin{eqnarray}
\Sigma(z)&=&\sum_{q}\sum_i P_q \Sigma_i^{(q)}(z)
\nonumber\\
&=&\sum_{q}\sum_i P_q v^{(q)}(z)|i><i|,
\label{ansatzS}
\end{eqnarray}
with $v^{(q)}(z)=v(z-q\Omega)$. From now on we shall adopt the 
convention that any operator or function
with a superscript $(q)$ has to be taken at the energy $z-q\Omega$.
Note,
in accordance with the spirit of the CPA, that the coherent potential 
$v(z)$ does not contain any spatial information: 
It neither depends on the site nor on the 
spatial distribution of the excited phonons.

Combining this ansatz for $\Sigma(z)$ with 
Eq. (\ref{geff}), we obtain for the effective one-resolvent   
\begin{eqnarray}
G_{eff}(z)=\sum_{q} P_q g^{(q)}(z),
\end{eqnarray}
with auxiliary (electronic) operators 
\begin{eqnarray}
g^{(q)}(z)=
{1\over{z-q\Omega-\sum_{\vec{k}}[\epsilon_B+\epsilon_{\vec{k}}+v^{(q)}(z)]|\vec{k}><\vec{k}|}}.
\label{gq}
\end{eqnarray}

Apparently, in contrast to the full one-resolvent $G(z)$, the effective one-resolvent
$G_{eff}(z)$ is a simple operator in the phonon Hilbert space. For an arbitrary phonon state  
$|\{n_l\})$ 
\begin{eqnarray}
G_{eff}(z)|\{n_l\})=g^{(\sum_l n_l)}(z)|\{n_l\})
\label{aux}
\end{eqnarray}
holds, i.e., (phonon) subspaces with different total phonon number are decoupled and, as a 
consequence,  
the self-consistency condition Eq. (\ref{sce1}) becomes
\begin{eqnarray}
\bar{G}(z)=g^{(0)}(z).
\label{sce2}
\end{eqnarray}

In order to derive a functional equation for the unknown 
function $v(z)$, which in turn specifies $\bar{G}(z)$, we perform a 
multiple-scattering analysis in the product Hilbert space 
of the coupled electron-phonon system.
Except for the underlying Hilbert space, the 
mathematical manipulations parallel the CPA multiple-scattering analysis of ref. 
\cite{V68}. The basic relation between the full and effective one-resolvents, 
\begin{eqnarray}
G(z)&=&G_{eff}(z)+G_{eff}(z)T(z)G_{eff}(z),
\label{res}
\end{eqnarray}
involves the total T-matrix, 
\begin{eqnarray}
T(z)&=&\sum_i Q_i(z),
\label{tmatrix}
\end{eqnarray}
which, according to ref. \cite{V68}, can be 
expressed in terms of individual single-site contributions  
\begin{eqnarray}
Q_i(z)&=&t_i(z)\left[ 1 + G_{eff}(z)\sum_{j\neq i} Q_j(z) \right],
\label{Q}
\end{eqnarray}
with the atomic T-matrix given by
\begin{eqnarray}
t_i(z)=\left[ 1-\Delta H_i(z) G_{eff}(z) \right]^{-1} \Delta H_i(z).
\label{tatom}
\end{eqnarray}
From Eq. (\ref{Q}) we see that $Q_i(z)$ is a product of the atomic T-matrices $t_i(z)$ 
and an effective wave factor 
$\left[ 1 + G_{eff}(z)\sum_{j\neq i} Q_j(z) \right]$
describing an effective wave incident on site $i$ modified by 
multiple-scattering events.
The single-site perturbation
\begin{eqnarray}
\Delta H_i(z)&=&
\left[\delta x_i - g\left(b_i^\dagger+b_i\right) - \sum_q P_q v^{(q)}(z)\right]
\nonumber\\
&\times& |i><i|
\end{eqnarray}
comprises bi-modal alloy type fluctuations and coupling to localized 
phonons on site $i$. 

Combining  
Eq. (\ref{res}) with Eq.(\ref{tmatrix}) and taking Eq. (\ref{aux}) into account,
the self-consistency condition Eq. (\ref{sce2}) is satisfied if we enforce
\begin{eqnarray}
\langle\langle(0|Q_i(z)|0)\rangle\rangle=0.
\label{sce3}
\end{eqnarray}
Iteration of Eq. (\ref{sce3}) gives rise to an intractable infinite series 
involving {\it all} sites of the lattice. The standard procedure to 
obtain a numerically feasible description is to factorize 
the average in  
Eq. (\ref{sce3}) into a product of averages, i.e., to 
average atomic T-matrices and effective wave factors separately.
Relegating a discussion of the factorization procedure to the 
next subsection, we approximate
Eq. (\ref{sce3}) by
\begin{eqnarray}
\langle\langle(0|Q_i(z)|0)\rangle\rangle&\approx&
\langle\langle(0|t_i(z)|0)\rangle\rangle
\nonumber\\
&\times& \langle\langle
(0|1 + G_{eff}(z)\sum_{j\neq i} Q_j(z)|0)\rangle\rangle,
\end{eqnarray}
reducing thereby the self-consistency condition to    
$\langle\langle(0|t_i(z)|0)\rangle\rangle=0$.
Because phonons on sites $j\neq i$
are not affected by scattering events encoded in $t_i(z)$  (they
are ``frozen'' and act only as ``spectators''), the 
single-site self-consistency condition can be formulated more
generally as
\begin{eqnarray}
\langle\langle(N,0_i|t_i(z)|0_i,N)\rangle\rangle&=&
\langle\langle(0,0_i|t_i^{(N)}(z)|0_i,0)\rangle\rangle 
\nonumber\\
&=& 0 ~~\forall~~N,
\label{sce5}
\end{eqnarray}
where $|n_i,N)$ denotes an arbitrary phonon state with $n$ phonons on site 
$i$ and 
a total number of $N$ phonons on all the other sites.  
Note the usage of the supersript notation defined above. The physical
interpretation of Eq. (\ref{sce5}) is the following:  
The electron ``remembers'' how many virtual phonons it has excited in 
previous scattering events (and, accordingly, how much energy is 
stored in the lattice); it does   
not ``remember'', however, at which sites these events took place.

Instead of working directly with Eq. (\ref{sce5}), 
it is more convenient to introduce an 
equivalent 
polaron-impurity model
(PIM)\cite{A88a}, which describes
a single perturbation $\Delta H_i(z)$ 
self-consistently embedded into an effective medium. 
Mathematically, the PIM is formulated by 
\begin{eqnarray}
D_i(z)&=&G_{eff}(z)+G_{eff}(z)t_i(z)G_{eff}(z)
\label{pim1}\\
&=&G_{eff}(z)+G_{eff}(z)\Delta H_i(z)D_i(z).
\label{pim2}
\end{eqnarray}
The self-consistency equation (\ref{sce5}) is then transformed into
\begin{eqnarray}
\langle\langle (0,0_i|D_i^{(N)}(z)|0_i,0) \rangle\rangle &=& g^{(N)}(z),
\end{eqnarray}
which in turn yields for the averaged one-resolvent
\begin{eqnarray}
\bar{G}(z)=g^{(0)}(z)=\langle\langle (0,0_i|D_i(z)|0_i,0) \rangle\rangle.
\label{sce6}
\end{eqnarray}

As shown in appendix A, it is straighforward to derive from the    
PIM a non-linear functional equation for
the local two-point function. We get
\begin{eqnarray}
&{\cal G}_{ii}&(z)=
\nonumber\\
&&\frac{1-c}{\displaystyle [F^{A,(0)}(z)]^{-1}-
   \frac{g^2}{\displaystyle [F^{A,(1)}(z)]^{-1}-
   \frac{2g^2}{\displaystyle [F^{A,(2)}(z)]^{-1}-
   }}}
\nonumber\\
&+&
   \frac{c}{\displaystyle [F^{B,(0)}(z)]^{-1}-
   \frac{g^2}{\displaystyle [F^{B,(1)}(z)]^{-1}-
   \frac{2g^2}{\displaystyle [F^{B,(2)}(z)]^{-1}-
   }}}
\label{cfraction}
\end{eqnarray}
with
\begin{eqnarray}
&F&^{\lambda,(n)}(z)=
\nonumber\\
& &{1 \over {z-n\Omega-\epsilon_\lambda
+{\cal G}_{ii}^{-1}(z-n\Omega)
-{\cal R}[{\cal G}_{ii}(z-n\Omega)]}},
\label{fnl2}
\end{eqnarray}
where ${\cal R}[\xi]$ denotes the inverse Hilbert 
transform (see below).

For $c=0$ 
Eqs. (\ref{cfraction})--(\ref{fnl2}) have been given before by 
Sumi \cite{S74}
and independently by Ciuchi et al. \cite{C97}. Various
limits of the $c=0$ equations, e.g., weak, intermediate, and 
strong electron-phonon coupling limits, 
have been discussed in these references.    
The non-linear functional couples local 
two-point functions with energies shifted by any {\it negative} 
integer multiple of $\Omega$. This accounts for the fact that
at zero temperature only phonon emission is possible. 
At finite temperature, in contrast, absorption of thermally 
excited phonons is also allowed. In that case,  
local two-point functions with energies shifted by {\it any}
integer multiple of $\Omega$, negative as well as 
positive, would be coupled \cite{C97,S74}.  

The coherent potential $v(z)$ can be directly obtained from Eq. 
(\ref{sce6}). Specifically, introducing the Hilbert transform 
${\cal H}[\xi]$ corresponding
to the bare density of states
$N_0(E)=(1/N)\sum_{\vec{k}} \delta(E-\epsilon_{\vec{k}})$, 
the local two-point function can be written as  
\begin{eqnarray}
{\cal G}_{ii}(z)
&=&\int dE {{N_0(E)} \over {z-\epsilon_B - v^{(0)}(z)-E} }
\nonumber\\
&=&{\cal H}[z-\epsilon_B-v^{(0)}(z)].
\label{Hilbert}
\end{eqnarray}     
Employing the inverse Hilbert transform, defined by ${\cal R}*{\cal H}[\xi]=\xi$,
finally leads to
\begin{eqnarray}
v^{(0)}(z)=z-\epsilon_B-{\cal R}[{\cal G}_{ii}(z)].
\label{self}
\end{eqnarray}
The local two-point function ${\cal G}_{ii}(z)$ uniquely determines the coherent 
potential $v^{(0)}(z)$. Once $v^{(0)}(z)$ is known any 
matrix element of the averaged one-resolvent can be obtained.

\subsection{Validity of the Factorization Procedure}

In the previous subsection we employed a factorization procedure
to reduce the multi-site self-consistency condition
to a single-site condition. Clearly, the 
mathematically exaxt average of the operator $Q_i$ contains many
terms which are not zero, even if the atomic T-matrix vanishes. 
The single-site self-consistency condition can be only
approximate. We now analyze the validity of the factorization
procedure taking advantage of the connection between the 
DCPA and the DMFA. As a by-product, we shall find an efficient way to derive
the DCPA equations for the averaged two-resolvent.

To that end, we follow ref. \cite{SE72} and rearrange the 
multiple-scattering series into clusters containing a fixed number
$p$ of lattice sites. We write (suppressing the energy variable $z$)
\begin{eqnarray}
Q_i=\sum_p Q_i^{[p]}
\end{eqnarray}
with $Q_i^{[1]}=t_i$ and 
\begin{eqnarray}
Q_i^{[p]}&=&t_i\sum_{j_1\neq i}G_{eff}t_{j_1}^{[2]}(i)\sum_{j_2\neq i,j_1}
G_{eff}t_{j_2}^{[3]}(i,j_1)
\nonumber\\
&...&\sum_{j_{p-1}\neq i,j_1,...,j_{p-2}} 
G_{eff}t_{j_{p-1}}^{[p]}(i,j_1,...,j_{p-2})
\label{Qcluster}
\end{eqnarray} 
for $p\ge 2$. The cluster T-matrices 
$t_{j_{p-1}}^{[p]}(i,j_1,...,j_{p-2})$ are defined as the 
sum of all scattering processes  
involving at most $p$ {\it different} sites  $i,j_1,...,j_{p-1}$
with the constraint that the entrance site has to be 
$j_{p-1}$. This constraint ensures that all p sites contributing
to $Q_i^{[p]}$ are connected.
The exit site, on the other hand, 
can be any site of the p-cluster.   
Notice that, in contrast to the original 
multiple-scattering series given in Eq. (\ref{Q}), where only
successive summation indices had to be different, {\it all} summation 
indices in Eq. (\ref{Qcluster}) are different.

\begin{figure}[t]
\hspace{2.0cm}\psfig{figure=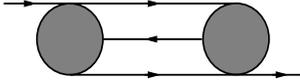,height=1.0cm,width=4.0cm,angle=-0}
\vspace{1.0cm}
\caption[fig1]
{Schematic representation of a fourth order process contributing 
to $Q_i^{[2]}$. Atomic T-matrices 
and effective one-resolvents are, respectively, depicted by solid circles
and solid lines. The arrows indicate the order in which the diagram has
to be read. The energy arguments are clear from the context. 
}
\label{fig1}
\end{figure}

We are now in a position to evaluate the importance of each 
individual multiple-scattering term contributing to $Q_i$. 
Keeping in mind that  
all processes where a site $j$ 
is visited only once vanish due to the single-site self-consistency 
condition Eq. $(\ref{sce5})$, only processes  
where {\it all} sites are {\it at least} visited twice
require further analysis. Figure \ref{fig1} schematically depicts 
the first non-vanishing $p=2$ process (i.e., a 
fourth order process in which each of
the two sites is visited twice). Writing 
\begin{eqnarray}
t_{nm}^{(r)}=<i|(0,n_i|t_i^{(r)}|m_i,0)|i> 
\label{tnm}
\end{eqnarray} 
for the (configuration dependent) matrix elements of the atomic T-matrix,
this process is analytically given by 
\begin{eqnarray}
Q_i^{[2],{4^{th}}}=
\sum_{j\neq i,qr}t_{0r}^{(0)}{\cal G}_{ij}^{(r)} 
t_{0q}^{(r)}{\cal G}_{ji}^{(r+q)} 
t_{r0}^{(q)}{\cal G}_{ij}^{(q)} 
t_{q0}^{(0)}
|i><j|.
\label{Qthird}
\end{eqnarray}

To proceed further, we now imagine at this point that the 
hopping matrix element is scaled 
by $1/\sqrt{Z}$ (as in the case of the DMFA). 
Then, we immediately see that for
large $Z$ Eq. (\ref{Qthird}) is at least $\sim 1/ \sqrt{Z}$ 
because it contains three off-diagonal two-point functions and only one
free summation over sites $j$, i.e., for $Z\rightarrow \infty$ 
this process does not contribute to $Q_i$ at all. In fact, 
it can be shown that all p-cluster processes  
in which {\it all} $p$ sites are at least visited twice are suppressed by 
a factor containing the inverse of the coordination number $Z$. 
As a consequence, for $Z\rightarrow \infty$ these processes vanish and need
not be explicitly considered in the averaging procedure.  

The implication of the above discussion is three-fold. 
First, it corroborates, 
using multiple-scattering theory, the conventional CPA technique, 
previous work by Vlaming and Vollhardt \cite{VV92}, who showed,
acknowledging earlier work by Schwartz and Siggia \cite{SS72}, that for a 
lattice with infinite coordination number any CPA-type theory becomes 
exact, i.e., for infinite coordination number, 
the single-site self-consistency
condition is indeed sufficient to ensure that the average of $Q_i$ as 
a whole vanishes.  

Second, it indicates that even for a finite coordination number, 
processes which do not
vanish due to the single-site self-consistency condition are 
at least suppressed.

Third, from a formal point of view, the DCPA is equivalent to   
the replacement of the full
operator $Q_i$ by a reduced operator $Q_i^{SAP}$ containing only 
{\it self-avoiding paths} (SAP);
$Q_i^{SAP}$ can be 
obtained from Eq. (\ref{Qcluster}) by setting
$t_{j_{p-1}}^{[p]}(i,j_1,...,j_{p-2})=t_{j_{p-1}} \forall~p$.  
The coherent potential $v(z)$ is then adjusted in such 
a way that the averaged $Q_i^{SAP}$ vanishes.
This procedure can be envisaged, in physical terms, as a 
{\it loss of spatial memory}, because the electron 
``assumes'' that each site it encounters is visited the first time, i.e.,
the electron does not ``remember'' whether it has visited the
site before. Naturally, for a lattice with infinite coordination
number, this is exact, because the probability that the electron
returns to the same lattice site is vanishingly small.

The replacement $Q_i \rightarrow Q_i^{SAP}$, i.e., the
erasur of the spatial memory of the electron, is of course 
consistent with the ansatz for the (local)
self-energy operator. On the other hand, this replacement 
leads to an insufficient treatment of inter-site processes 
in the calculation of the averaged two-resolvent and, eventually, 
prevents a rigorous calculation of the
diffusion and trapping properties. 
 
Keeping this shortcoming in mind, we shall nevertheless employ 
this formal replacement
from the outset in the next subsection to  
simplify the
calculation of the four-point function. 

\subsection{Calculation of $\bar{K}(z_1,O_e,z_2)$}

The discussion of the previous subsection 
suggests that the single-site self-consistency condition 
is, from a formal point of view,
equivalent to the replacement 
$Q_i \rightarrow Q_i^{SAP}$, i.e., only SAPs need to be 
explicitly considered in the averaging procedure.
We now show that this replacement allows a rather straightforward 
derivation of the averaged two-resolvent. 

For that purpose, we 
first recast Eq. (\ref{kbar}), using
Eqs. (\ref{aux}) and (\ref{res}) together with
the self-consistency condition,
Eq. (\ref{sce5}), into 
\begin{eqnarray}
\bar{K}(z_1,O_e,z_2)&=&
g^{(0)}(z_1)\left[O_e + \bar{\Gamma}(z_1,O_e,z_2)\right]
\nonumber\\
&\times& g^{(0)}(z_2).
\label{kr1}
\end{eqnarray}
Keeping only site-diagonal 
terms, the electronic vertex operator is given by 
\begin{eqnarray}
\bar{\Gamma}(z_1,O_e,z_2)=\sum_i
\langle\langle(0|\Gamma_i(z_1,O_e,z_2)|0)\rangle\rangle,
\label{vertex1}
\end{eqnarray}
with a local vertex operator
\begin{eqnarray}
&\Gamma&_i(z_1,O_e,z_2)=
\nonumber\\
&Q_i&^{SAP}(z_1)
G_{eff}(z_1)O_eG_{eff}(z_2)\tilde{Q}_i^{SAP}(z_2).
\label{vertex2}
\end{eqnarray}

\begin{figure}[t]
\hspace{2.0cm}\psfig{figure=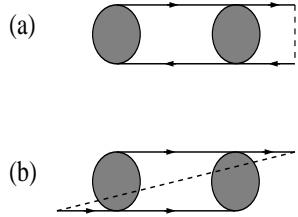,height=3.0cm,width=4.0cm,angle=-0}
\vspace{1.0cm}
\caption[fig2]
{Schematic representation of the second order (ladder) process
contributing to the local DCPA vertex (a) and of the
second order (maximally crossed) process neglected within the DCPA (b).
The dashed line depicts
the operator $O_e$ and solid circles and lines stand, respectively, for
atomic T-matrices and effective one-resolvents.
The arrows indicate the order in which the diagrams have to
be read. The energy arguments are clear from the context.
}
\label{fig2}
\end{figure}

In the above equations we have kept only
retraceable SAP's where 
the second leg of the path, encoded in $\tilde{Q}_i^{SAP}$,
retraces in {\it reversed} order the first leg of the path, 
encoded in $Q_i^{SAP}$ and neglected retraceable SAP's where the 
second leg retraces the first leg in the {\it same} order.
In more familiar terms (see Fig. \ref{fig2}), we have kept the ladder diagrams and
neglected (maximally) crossed diagrams. This choice is consistent 
with the coherent potential determined in the previous subsection.
In particular, it can be shown that
the local vertex defined in the ladder approximation and the local self-energy
obey the Ward
identity corresponding to particle conservation \cite{V69}. 

In analogy with the 
classical alloy problem, we expect the maximally crossed diagrams
to yield a significant (for certain model parameters possibly diverging) 
contribution to the local vertex function. As a consequence, a 
consistent treatment of these processes might render the potential to
force $\lim_{\eta\rightarrow 0}P(\omega,\eta)$ to be finite for 
some energies $\omega$ and for certain model parameters. 
Accordingly, itinerant and trapped polaron states
could be rigorously distinguished. We did not, however, succeed in 
taking them into account. 

\begin{figure}[t]
\hspace{2.0cm}\psfig{figure=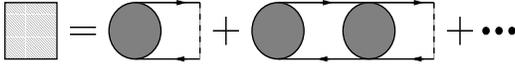,height=1.0cm,width=7.0cm,angle=-0}
\vspace{1.0cm}
\caption[fig3]
{Schematic representation of Eq. (\ref{vertex2}). The box on the left  
hand side denotes the local vertex operator $\Gamma_i$, the dashed line depicts
the operator $O_e$, and solid circles and lines stand, respectively, for
atomic T-matrices and effective one-resolvents.
The arrows indicate the order in which the diagram has to
be read. The energy arguments are clear from the context.
}
\label{fig3}
\end{figure}

To derive an explicit equation for the local vertex function (in
the ladder approximation), we have to
perform the average of Eq. (\ref{vertex2}). To that end, it is advantageous
to derive a formal operator equation for the local vertex operator $\Gamma_i$. 
For simplicity, we shall from now on suppress the arguments of the
various operators.
The implied arguments should be clear from the context.       
If we  
visualize 
Eq. (\ref{vertex2}) in terms of diagrams (see Fig. \ref{fig3}), 
we readily obtain 
\begin{eqnarray}
\Gamma_i=t_i G_{eff}\left[O_e + \sum_{j\neq i}\Gamma_j^{(-i)}\right] 
G_{eff}t_i, 
\label{vertex3}
\end{eqnarray}
where $\Gamma_j^{(-i)}$ stands for the vertex operator on site $j$ 
given by summing up all ``ladder-type paths'' not visiting site $i$. 
Consequently, as far as the configuration average 
is concerned, the atomic T-matrices $t_i$ and the term between the brackets 
are statistically independent and the 
configuration average factorizes, i.e.,
\begin{eqnarray}
\langle\langle \Gamma_i \rangle\rangle=
\langle\langle t_i G_{eff}
\langle\langle\left[O_e + \sum_{j\neq i}\Gamma_j^{(-i)}\right]\rangle\rangle
G_{eff} t_i \rangle\rangle.
\label{vertex4}
\end{eqnarray}                            

For a macroscopic solid we expect
$\langle\langle\Gamma_j^{(-i)}\rangle\rangle$ essentially to be 
the same operator as $\langle\langle\Gamma_j\rangle\rangle$. 
Hence, if we dealt only with alloy-type disorder, Eq. (\ref{vertex4})
would be the result given by 
Velick\`y \cite{V69}. 
To cope with inelastic scattering, we have to keep in mind, however,
that the vertex function (similar to the coherent potential) depends
only on the electron energy, i.e., we have to keep track of the 
energy stored in the lattice (given by the total number of virtual phonons).
Consistent with the ansatz for the 
local self-energy operator, we therefore write
\begin{eqnarray}
\langle\langle\Gamma_i\rangle\rangle&=&
\sum_q P_q \Gamma_i^{(q)}
\label{ansatzG1}
\nonumber\\
&=&\sum_q P_q \gamma_i^{(q)}|i><i|,
\label{ansatzG}
\end{eqnarray}
with $\gamma_i^{(q)}=\gamma_i(z_1-q\Omega,O_e,z_2-q\Omega)$. Thus, 
the phonon vacuum expectation value of Eq. (\ref{vertex4}) becomes
\begin{eqnarray}
\gamma_i^{(0)}=
\langle\langle\sum_q t_{0q}^{(0)}
\left[K_i^{(q)}-{\cal G}_{ii}^{(q)}\gamma_i^{(q)}
{\cal G}_{ii}^{(q)}\right]
t_{q0}^{(0)} \rangle\rangle,
\label{vertex5}
\end{eqnarray}
with 
\begin{eqnarray}
K_i^{(q)}=<i|g^{(q)}O_eg^{(q)}|i>
+\sum_l{\cal G}_{il}^{(q)}\gamma_l^{(q)}{\cal G}_{li}^{(q)}.
\label{kr2}
\end{eqnarray}
Note, the only 
configuration dependent quantities are now the matrix elements of the
atomic T-matrix. 

The linear functional for the vertex function is not ``local''
in the energy variables $z_1$ and $z_2$. Instead, due to 
the possibility of phonon emission,
Eq. (\ref{vertex5}) 
couples all vertex functions with energies shifted by any {\it negative}
integer multiple of $\Omega$. 

Employing spatial Fourier transforms,
\begin{eqnarray}
K_{\vec{Q}}^{(r)}&=&\sum_i e^{-i\vec{Q}\cdot\vec{R}_i} K_i^{(r)},
\\
a_{\vec{Q}}^{(r)}&=&\sum_i e^{-i\vec{Q}\cdot\vec{R}_i} <i|g^{(r)}O_eg^{(r)}|i>,
\\
{\cal A}_{\vec{Q}}^{(r)}&=&\sum_{\vec{R}_i-\vec{R}_j} 
e^{-i\vec{Q}\cdot [\vec{R}_i-\vec{R}_j]}
{\cal G}_{ij}^{(r)}{\cal G}_{ji}^{(r)},
\label{aqr}
\\
\gamma_{\vec{Q}}^{(r)}&=&\sum_i e^{-i\vec{Q}\cdot\vec{R}_i} \gamma_i^{(r)},
\end{eqnarray}
it is straigthforward to combine Eqs. (\ref{vertex5}) and (\ref{kr2}) 
and to derive, after 
some rearrangements, 
a single equation for the Fourier-transformed local vertex function:
\begin{eqnarray}
\gamma_{\vec{Q}}^{(0)}&=&
\sum_q\langle\langle t_{0q}^{(0)}t_{q0}^{(0)} \rangle\rangle a_{\vec{Q}}^{(q)}
\nonumber\\
&+&\sum_q\langle\langle t_{0q}^{(0)}t_{q0}^{(0)} \rangle\rangle
\{ {\cal A}_{\vec{Q}}^{(q)}-{\cal G}_{ii}^{(q)}{\cal G}_{ii}^{(q)}\}
\gamma_{\vec{Q}}^{(q)}.
\label{vertex6}
\end{eqnarray}

This equation reduces in the limit $\Omega=0$
to Velick\'y's vertex equation \cite{V69}
for a model with a
Gaussian distribution of on-site energies on top of a bi-modal
distribution of on-site
energies.       
Note, moreover, that $\gamma_{\vec{Q}}^{(0)}$ 
may be used to calculate {\it any}
matrix element of the averaged two-resolvent of interest, i.e.,
the averaged two-resolvent is completely determined by $\gamma_{\vec{Q}}^{(0)}$.  

\section{Numerical Results for the $Z=\infty$ Bethe Lattice}

\subsection{Final Set of Equations}

In the previous section, we neither specified the coordination number
$Z$ nor the structure of the underlying lattice.
The DCPA equations for the one- and two-resolvents can be used for
arbitrary lattices. On the other hand, we have shown in 
subsection II.C that the single-site
factorization procedure essentially ignores processes, 
where the electron returns to the same lattice site. It is therefore
consistent with the DCPA to use a lattice, which, by construction,
completely prohibits this type of processes. A particularly simple case 
is the $Z=\infty$ Bethe lattice. (It should be emphasized, however,   
that, within the DCPA, the numerical results, in particular for the 
spectrally resolved return probability, are not affected much by
the chosen lattice.)

The main simplification, as far as the calculation of the 
local two-point function 
is concerned, arises from the fact that the inverse 
Hilbert transform
for a $Z=\infty$ Bethe lattice is an algebraic relation, namely 
${\cal R}[\xi]=1/4\xi + 1/\xi$, which makes the self-consistent 
solution of Eqs. (\ref{cfraction})--(\ref{fnl2}) particularly easy. 

The calculation of the local vertex function 
simplifies even more, because, for a 
$Z=\infty$ lattice, Eq. (\ref{aqr}) 
reduces (for almost all $\vec{Q}$) to
${\cal A}^{(r)}_{\vec{Q}}(z_1,z_2)\approx{\cal G}_{ii}^{(r)}(z_1)
{\cal G}_{ii}^{(r)}(z_2)$ \cite{G96}. Therefore,
the second term on the rhs of Eq. (\ref{vertex6}) vanishes 
and the local vertex function becomes
\begin{eqnarray}
\gamma_{\vec{Q}}^{(0)}=
\sum_q\langle\langle t_{0q}^{(0)}t_{q0}^{(0)}\rangle\rangle a_{\vec{Q}}^{(q)},
\label{vertex7}
\end{eqnarray}
which yields for the Fourier transform of Eq. (\ref{kr2}):
\begin{eqnarray}
K_{\vec{Q}}^{(0)}=a_{\vec{Q}}^{(0)}+
{\cal G}_{ii}^{(0)}{\cal G}_{ii}^{(0)}\sum_q
\langle\langle t_{0q}^{(0)}t_{q0}^{(0)}\rangle\rangle a_{\vec{Q}}^{(q)}.
\label{KQvector}
\end{eqnarray}

To investigate the trapping properties of the polaron states, 
we need the four-point function 
$f(z_1,z_2)$ defined in Eq. (\ref{f1}). To that end, we
set $O_e=|i><i|$ in the above equation and obtain after
Fourier transformation
\begin{eqnarray}
f(z_1,&z_2&)=
{\cal G}_{ii}(z_1){\cal G}_{ii}(z_2)
+{\cal G}_{ii}(z_1){\cal G}_{ii}(z_2)
\nonumber\\
&\times&\sum_q \langle\langle t_{0q}^{(0)}(z_1)t_{q0}^{(0)}(z_2)\rangle\rangle
{\cal G}^{(q)}_{ii}(z_1){\cal G}^{(q)}_{ii}(z_2).
\label{f2}
\end{eqnarray}
This equation holds for any
lattice with $Z=\infty$. The structure of the lattice, i.e., in 
our case, the structure of the Bethe lattice, enters 
only through the local two-point function ${\cal G}_{ii}(z)$.

In appendix B we
show that Eq. (\ref{f2}) can be brought into a form which allows us to calculate 
$f(\omega-i\eta,\omega+i\eta)$ from the amplitudes 
${\cal D}_{r0}^{\lambda,(0)}(\omega+i\eta)$ defined in appendix A. 
Specifically,
\begin{eqnarray}
f(\omega-i\eta,\omega+i\eta)&=&(1-c)\sum_{r=0}^\infty
|{\cal D}_{r0}^{A,(0)}(\omega+i\eta)|^2 
\nonumber\\
&+&c\sum_{r=0}^\infty|{\cal D}_{r0}^{B,(0)}(\omega+i\eta)|^2,
\label{F3}
\end{eqnarray}
i.e., the calculation of the four-point function 
$f(\omega-i\eta,\omega+i\eta)$ requires 
the same (moderate) numerical effort as the calculation of the local 
two-point function 
${\cal G}_{ii}(\omega+i\eta)$. 

The numerical strategy is now as follows. First, we truncate 
the continued fraction expansions 
in Eq. (\ref{cfraction}) 
and find the fixed point ${\cal G}^{(fix)}_{ii}(\omega+i\eta)$  
by iteration. 
The density of states then follows directly from
\begin{eqnarray}
N(\omega)=-{1\over \pi} Im{\cal G}^{(fix)}_{ii}(\omega+i\eta).
\label{DOS}
\end{eqnarray}
To study the individual contributions of the A and B sites, we also
calculate the component density of states 
\begin{eqnarray}
N_\lambda(\omega)=-{1\over \pi} Im {\cal D}^{\lambda,(0)}_{00}(\omega+i\eta).
\label{compDOS}
\end{eqnarray}
The amplitudes ${\cal D}^{\lambda,(0)}_{00}$, which are also
needed in the calculation of the spectrally resolved return 
probability (see below),
are given by the first ($\lambda=A$) and the second 
($\lambda=B$) term on the rhs of
Eq. (\ref{cfraction}) without the factors $1-c$ and $c$, respectively,
and with ${\cal G}_{ii}(z)={\cal G}^{(fix)}_{ii}(z)$.           

Second, we insert ${\cal G}^{(fix)}_{ii}(\omega+i\eta)$ in Eq. (\ref{self}) 
and obtain the coherent potential $v^{(0)}(\omega+i\eta)$, which yields, 
using Eqs. (\ref{gq}) and (\ref{sce2}), the spectral function
\begin{eqnarray}
A(\omega,\epsilon_{\vec{k}})=-{1\over \pi}
Im{1\over{\omega+i\eta-\epsilon_B-\epsilon_{\vec{k}}-v^{(0)}(\omega+i\eta)}}.
\label{SF}
\end{eqnarray}   

Third, to calculate the spectrally resolved return probability 
$P(\omega,\eta)$ defined in Eq. (\ref{returnprob2}), 
we truncate the sums in 
Eq. (\ref{F3}) and successively construct the
amplitudes ${\cal D}_{r0}^{\lambda,(0)}(\omega+i\eta)$ 
from the recursion relation, Eq. (\ref{RR}), starting from 
${\cal D}_{00}^{\lambda,(0)}(\omega+i\eta)$ with $\lambda=A,B$.
The truncation procedures and the subsequent numerical solution of the truncated 
equations for small $\eta>0$ converge very well.

\subsection{Spectral properties}

In this subsection we discuss the spectral properties 
of a subclass of polaron states, which we expect to be very susceptible 
to an (intrinsic or extrinsic) trapping transition. However, as mentioned
in the Introduction, spectral properties alone cannot be used, 
even within a rigorous theory, 
to decide whether states are itinerant or trapped.  

We start with the
ordered Holstein model [model (\ref{model}) with $\delta=0$], which 
is characterized by two parameters, 
$\lambda=2g^2/\Omega$ and $\alpha=g/\Omega$. (Recall, we measure 
energy in units of $2J$.) These  
parameters are conventionally used to define 
strong, intermediate, and weak electron-phonon coupling regimes 
\cite{F90}. The overall
DCPA spectral features of which have been 
thoroughly investigated by Ciuchi et al. \cite{C97} and by Sumi \cite{S74}. 
The most intriguing of these regimes is, perhaps, the intermediate 
electron-phonon coupling 
regime, approximately defined by   
$\lambda \ge 1$ and $\alpha \ge 1$, where a finite number of polaronic
subbands starts to develop on the low energy side of the bare 
density of states. The formation of these polaronic subbands, in particular
as a function of $\lambda$ and $\alpha$, has been intensively studied 
in above mentioned references. For our purpose, it is 
sufficient to focus, for a fixed $\lambda$ and $\alpha$, on the 
{\it lowest} polaronic subband.  

Generic numerical results for the ordered Holstein model in the 
intermediate electron-phonon coupling regime are summarized
in Fig. \ref{fig4}. The
polaron parameters are given in Table \ref{table}. 
We first focus on panels (a) and (b), which depict 
the density of states $N(\omega)$, the imaginary part
of the coherent potential $Imv(\omega)$, and the spectral function 
$A(\omega,\epsilon_{\vec{k}})$, respectively. The 
spectrally resolved return probability $P(\omega,\eta)$ shown in 
panel (c) will be discussed in the next subsection. 

The main spectral characteristics in the intermediate electron-phonon
coupling regime is the formation of a well-established lowest  
polaronic subband with an asymmetric, steeple-like density 
of states, separated from the rest of the spectrum
by a hard gap. Below the phonon emission threshold, which is inside
the second polaronic subband, no residual scattering 
takes place, i.e., $Imv(\omega)=0$. 
That is, within the lowest 
polaronic subband,
the electron and the phonons 
arrange themselves into a new (composite) entity -- a polaronic
quasi-particle \cite{C97}.
[The pole of $Imv(\omega)$ slightly above the
high-energy edge of the lowest polaronic subband does not imply a
diverging scattering rate; it merely signals the presence of the
gap.]
That the {\it quasi-particle concept} indeed 
applies can be 
also seen from the spectral function, which features a  
(dispersive) quasi-particle peak in the 
energy range of the lowest polaronic subband.
Notice the pronounced band 
flattening and the 
vanishing quasi-particle weight in the flat part of the 
dispersion. The band flattening is the reason for the 
steeple-like density of states. At higher energies
a second quasi-particle peak corresponding to the 
second polaronic subband is visible until the   
point where it merges 
with the one-phonon scattering continuum.  

Both the steeple-like shape of the density of states and the band
flattening at the high-energy edge of the lowest polaronic subband are an
{\it indirect}
consequence of the pole in the imaginary part of the coherent potential.
As can be seen in Figs. \ref{fig5}a and \ref{fig5}b, the pole in the imaginary
part is accompanied by a divergence in the real part. [Both are 
broadened due to the finite value of $\eta$; the 
pole becomes a Lorentzian and the divergence 
becomes a pronounced kink.]
The divergence, in turn, induces a pronounced energy dependence of 
$Rev(\omega)$ just below the high-energy edge and it is this 
strong energy dependence which, as we shall show below,
eventually yields the steeple-like 
density of states and the band flattening. Physically, the energy
dependence of the real part of the coherent potential reflects the
increasing phonon admixture in the states as the
high-energy edge of the subband is approached.

The discussion of the steeple-like subband density of states is
facilitated by considering  
\begin{eqnarray}
N_{1^{st}}(\omega)={2\over\pi}Re\sqrt{1-[\omega-\epsilon_B-Rev(\omega)]^2},
\label{subDOS}
\end{eqnarray}
an exact analytical expression for the lowest subband density
of states,
which follows from Eqs. (\ref{DOS}) and (\ref{Hilbert}) in the 
limit $\eta\rightarrow 0$ and the fact that for the lowest subband 
$Imv(\omega)=0$. Two conclusions can be drawn from Eq. (\ref{subDOS}).
First, the position $\omega_m$ of the
maximum of the subband density of states does not coincide with the
high-energy edge $\omega_h$, because they are respectively given by 
$\omega_m-\epsilon_B-Rev(\omega_m)=0$ 
(thin solid line in Fig. \ref{fig5}a) and
$\omega_h-\epsilon_B-Rev(\omega_h)=1$
(thin dashed line in Fig. \ref{fig5}a). 
Hence, the density of states
at the high-energy egde is, albeit very step-like, 
not discontinuous, even for $\eta\rightarrow 0$.  
Second, Eq. (\ref{subDOS}) together with 
Figs. \ref{fig5}a and \ref{fig5}b reveal that the rapid change of 
$Rev(\omega)$ between
$\omega_m$ and $\omega_h$, caused by the divergence {\it above} $\omega_h$,
produces the rapidly dropping density of states in this 
particular energy range, and hence, the steeple-like shape.

Figures \ref{fig5}a and \ref{fig5}c suggest, moreover, that  
the rapidly changing $Rev(\omega)$ 
is also responsible for the band flattening.
To be more specific, we recall
that the dispersion $\omega(\epsilon_{\vec{k}})$ is given by the poles of the
spectral function defined in Eq. (\ref{SF}).
Therefore, the change of the dispersion with 
$\epsilon_{\vec{k}}$ becomes
\begin{eqnarray}
{{d\omega(\epsilon_{\vec{k}})}\over{d\epsilon_{\vec{k}}}}=
(1-{{dRev(\omega)}\over{d\omega}}|_{\omega=\omega(\epsilon_{\vec{k}})})^{-1},
\label{flat}
\end{eqnarray}
which is, of course, particularly small for $\omega_m\le\omega\le\omega_h$, 
since the divergence yields a large derivative of $Rev(\omega)$.  
It is, therefore, the {\it divergence-induced}  
strong $\omega$-dependence of the real part of the coherent potential
which not only causes the steeple-like subband density of states but also 
the pronounced flattening of the subband dispersion.

The DCPA results for the lowest polaronic subband corroborate recently 
obtained direct numerical simulation results for a finite 
Holstein model in the single-electron sector \cite{S96,WF97,BTB99}. 
Specifically, the 
band flattening
and the vanishing  
quasi-particle weight in the flat part of the dispersion
are in good qualitative agreement.   
Both effects, characteristic of the intermediate 
electron-phonon coupling
regime, are manifestations of the strong on-site 
electron-phonon correlations,
which yield an increasing 
phonon admixture in the states comprising the high-energy edge 
of the lowest
polaronic subband \cite{WF97}.  

Clearly, the phonon admixture in the states
must significantly affect the diffusion properties.
In particular, because the dispersion 
becomes extremely flat in the vicinity of the high-energy edge of the
lowest polaronic subband,  
these states might be 
{\it temporarily} trapped. In the 
next subsection, we will present an 
analysis of the spectrally resolved return probability $P(\omega,\eta)$,
which indeed shows that the high-energy edge states are extremely 
``sluggish''.

However, first we illustrate the modifications of the 
lowest polaronic subband due to alloying.
We expect disorder on the
scale of the subband width to show the most dramatic
effects and therefore restrict $\delta$ 
[in Eq. (\ref{model})] to small values. 
Further, to be specific, we chose the B atoms to 
be energetically close to the high-energy edge of the subbands, 
i.e., we consider $\delta<0$.

Figure \ref{fig6} shows representative numerical results for 
the Holstein alloy.
The polaron parameters are the same as in Fig. \ref{fig4}
(see Table \ref{table}) and the alloy parameters are
$c=0.5$ and $\delta=-0.01$. First, we focus again on panels 
(a) and (b), which summarize the spectral properties in 
the vicinity of the lowest polaronic subband. The 
spectrally resolved return probability shown in panel (c) 
will be discussed in the next subsection.

The main effect of alloying is 
the formation of two mini-subbands. 
For the chosen parameters both mini-subband density 
of states exhibit a pronounced asymmetric, steeple-like shape
(see Fig. \ref{fig6}a). 
From the strong asymmetry of the component density of states, 
which are also depicted in Fig. \ref{fig6}a,
we find, moreover, that, although both A- and B-sites contribute 
almost equally to both mini-subbands,
the steeple-like shape of the mini-subband density of states
comes entirely either from the A-sites 
(lower mini-subband) or from the B-sites (upper mini-subband).  
As in the case without disorder, the steeple-like shape is due to 
strong on-site
electron-phonon correlations, which yield a large phonon admixture
in the states at the high-energy edges of the respective mini-subbands, 
and,
as a consequence, to rather flat dispersions in these parts of
the spectrum. The band flattening can be clearly seen from the
spectral function shown in Fig. \ref{fig6}b. 

The detailed appearance of the mini-subbands depends on 
the concentration $c$ and the scattering strength $\delta$. 
To indicate the concentration 
dependence, for example, we show in Fig. \ref{fig7}
the density of states $N(\omega)$ for 
a fixed scattering strength $\delta=-0.004$ and 
concentrations varying
from the pure A crystal ($c=0$) to
the pure B crystal limit ($c=1$). Note, first, the
noticeable distortions of the density of states (at the
high-energy edge) despite the fact that $\delta$ is
much smaller than the polaronic subband width and, second, the lack of 
$c \leftrightarrow 1-c$ symmetry. That rather small amounts of 
disorder suffice to restructure the high-energy edge can be
also seen in Fig. \ref{fig8}, where the lower panel shows the 
density of states and the imaginary part of the coherent potential
for $c=0.5$ and three different  
scattering strengths $\delta$. 

It is not necessary to present an exhaustive investigation of
the alloying-induced spectral changes in the whole 
$c-\delta$ parameter plane. The data presented
suffice already to show that the modifications of the  
lowest polaronic subband are very similar to the ones found 
in an AB alloy with an {\it artificial} steeple-like bare density of 
states \cite{KVE70}. Indeed, 
the broken $c \leftrightarrow 1-c$ symmetry, the
appearance of a B mini-subband at scattering strengths
much smaller than the polaronic subband width, and the 
asymmetry between the component density of states have their
counterparts in this artificial steeple alloy. 

In the artificial steeple alloy, as well as in the Holstein alloy, 
states comprising the steeple-like structure in the density of
states are very susceptible to impurity scattering. For the
Holstein alloy this is intuitively clear, because the steeple
in the density of states contains the states corresponding
to the flat part of the
dispersion. As indicated above,   
due to the large phonon admixture, these states might be
temporarily trapped, which suggests
that they feel the on-site, alloy-type
disorder much more strongly than the other states. The discussion
of the diffusion properties in the next subsection will substantiate 
this intuitive picture. 

\subsection{Diffusion properties}

The spectral properties in the vicinity of the 
high-energy edge of the lowest polaronic subband suggest 
that this part of the spectrum is  
very susceptible to 
an (intrinsic or extrinsic) trapping transition. 

In the Introduction, we emphasized that the 
problem of trapping of a polaron should be 
methodologically addressed with the same techniques 
used to investigate the problem of localization of an electron 
in a random potential. 
In particular, we stressed that 
$\lim_{\eta\rightarrow 0} P(\omega,\eta)$ would be
a rigorous criterion 
to distinguish itinerant polaronic quasi-particle states
from (any) trapped polaronic defect states. We also 
pointed out, that a rigorous investigation of 
the trapping issue is unfortunately beyond the DCPA because, 
within the DCPA, $\lim_{\eta\rightarrow 0} P(\omega,\eta)$ 
vanishes for all energies $\omega$. Although trapped states cannot be unambiguously 
identified, it is possible to
identify {\it temporarily} trapped states through the 
$\eta$-asymptotics of $P(\omega,\eta)$.  

To motivate the investigation of the $\eta$-asymptotics of $P(\omega,\eta)$,
we first emphasize that the
spectrally resolved return 
probability $P(\omega,\eta)$  
contains valuable information 
about the diffusion properties not only in the limit $\eta=0$ 
(which is usually considered \cite{EC72}) but also 
for a finite $\eta>0$. 
For a correct interpretation of the data, we have
to keep in mind, however, that a finite $\eta$ broadens all spectral features.
In particular, the pole [divergence] in $Imv(\omega)$ [$Rev(\omega)$] 
becomes, as mentioned in the previous subsection, a Lorentzian 
[pronounced kink], smearing out the high-energy edges of the
respective bands. A finite $\eta$ yields therefore small
artificial tails extending into the gap regions.
These tails have no direct physical meaning. In the following we analyze,
therefore, the $\eta-$asymptotics of $P(\omega,\eta)$ only
for energies $\omega$ below the respective sharp high-energy edges obtained for
$\eta\rightarrow 0$ (see Fig. \ref{fig5}b and the insets of Figs. 
\ref{fig9}--\ref{fig11}). 

\begin{figure}[t]
\psfig{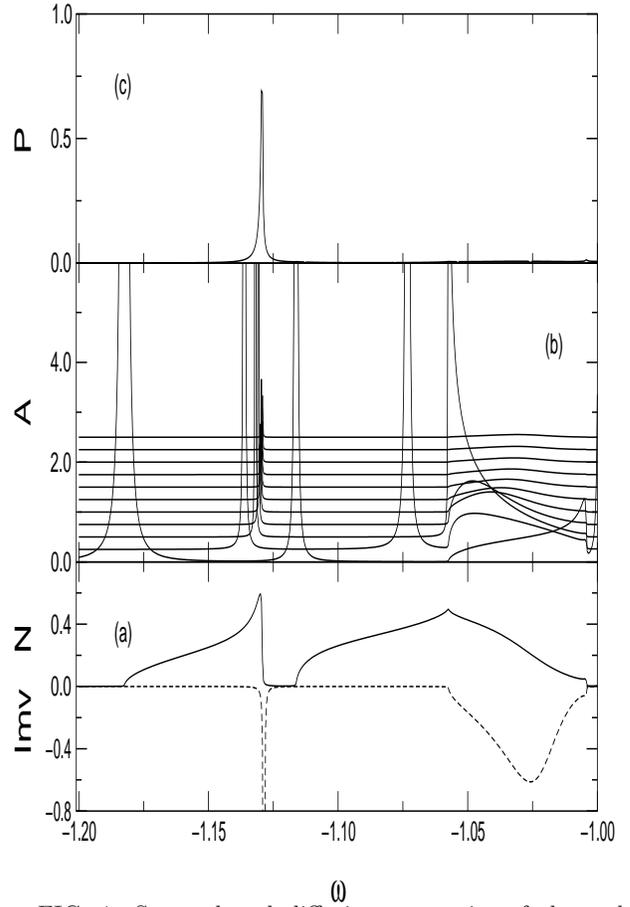}
\caption[fig4]
{Spectral and diffusion 
properties of the {\it ordered} Holstein model in the 
vicinity of the first polaronic subband for 
$\lambda=1.96$, $\alpha=2.8$ [see Table
\ref{table}], and $\eta=10^{-4}$.
(a) Density of states $N(\omega)$ (solid line) and imaginary part of
the coherent potential $Imv(\omega)$ (dashed line).           
(b) Spectral function $A(\omega,\epsilon_{\vec{k}})$ with $\epsilon_{\vec{k}}$
ranging from $\epsilon_{\vec{k}}=-1$ to $\epsilon_{\vec{k}}=1$
(bottom to top) in steps of $\Delta \epsilon_{\vec{k}}=0.2$. For
clarity the spectral functions corresponding to different
$\epsilon_{\vec{k}}$ are artificially shifted along the vertical axis.   
(c) Spectrally resolved return probability $P(\omega,\eta)$.
}
\label{fig4}
\end{figure}

\begin{figure}[t]
\psfig{figure=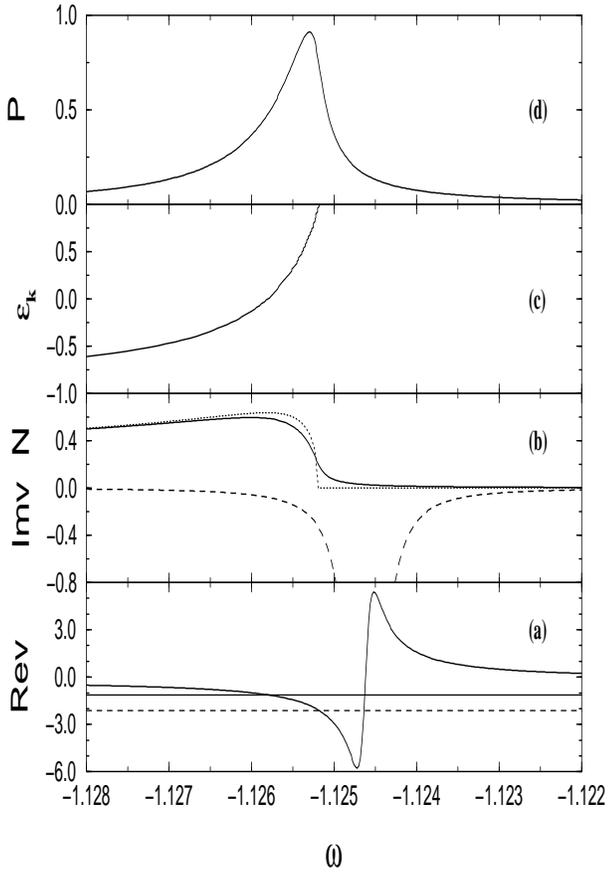,height=12.0cm,width=8.0cm,angle=-0}
\caption[fig5]
{Blow-up of Fig. \ref{fig4} in the immediate vicinity of 
the high-energy edge of the lowest polaronic subband.
(a) Real part of the coherent potential $Rev(\omega)$ for 
$\eta=10^{-4}$ (thick solid line). The 
intersections (below $\omega=-1.125$) with the thin solid and dashed lines 
determine, respectively, the position of 
the maximum of the subband density of states ($\omega_m$) and the 
position of the high-energy edge ($\omega_h$). 
(b) Density of states for $\eta=10^{-4}$ (solid line) and $\eta=10^{-8}$
(dotted line). The dashed line depicts the imaginary part of the 
coherent potential $Imv(\omega)$ for $\eta=10^{-4}$. 
(c) The poles of the spectral function $\omega(\epsilon_{\vec{k}})$. Here
we plot $\epsilon_{\vec{k}}$ vs. $\omega$. 
(d) Spectrally resolved return probability $P(\omega,\eta)$ for
$\eta=10^{-4}$.
}
\label{fig5}
\end{figure}

\begin{figure}[t]
\psfig{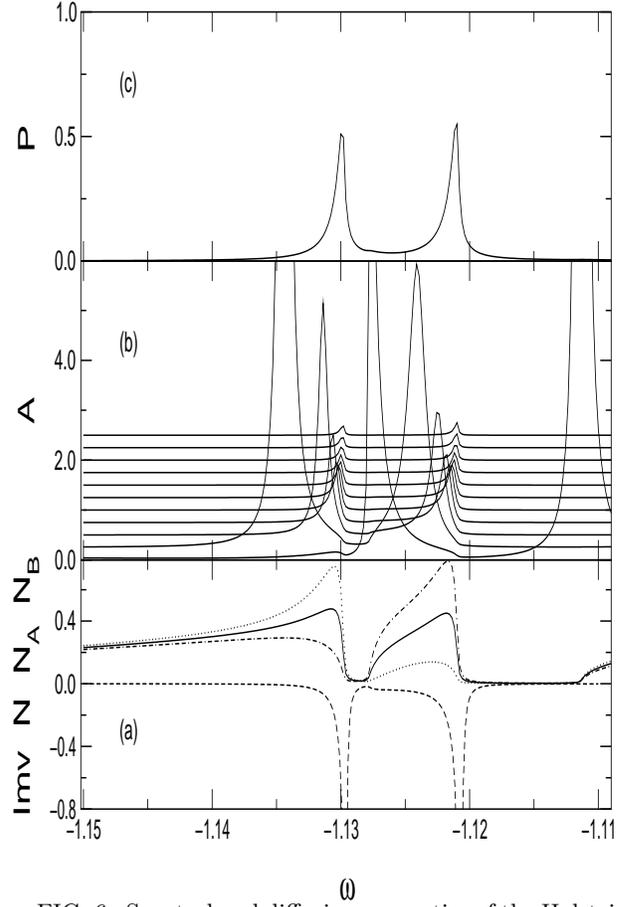}
\caption[fig6]
{Spectral and diffusion
properties of the Holstein alloy in the
vicinity of the high-energy edge of the
first polaronic subband for $c=0.5$ and
$\delta=-0.01$. The polaron parameters are
$\lambda=1.96$ and $\alpha=2.8$ [see Table
\ref{table}] and $\eta=10^{-4}$.
(a) Density of states $N(\omega)$ (solid line), component density
of states $N_A(\omega)$ (dotted line) and $N_B(\omega)$ (dot-dashed line),
and imaginary part of the coherent potential $Imv(\omega)$ (dashed line).     
(b) Spectral function $A(\omega,\epsilon_{\vec{k}})$ with $\epsilon_{\vec{k}}$
ranging from $\epsilon_{\vec{k}}=-1$ to $\epsilon_{\vec{k}}=1$
(bottom to top) in steps of $\Delta \epsilon_{\vec{k}}=0.2$. For
clarity the spectral functions corresponding to different
$\epsilon_{\vec{k}}$ are artificially shifted along the vertical axis.
(c) Spectrally resolved return probability $P(\omega,\eta)$.
}
\label{fig6}
\end{figure}

\begin{figure}[t]
\psfig{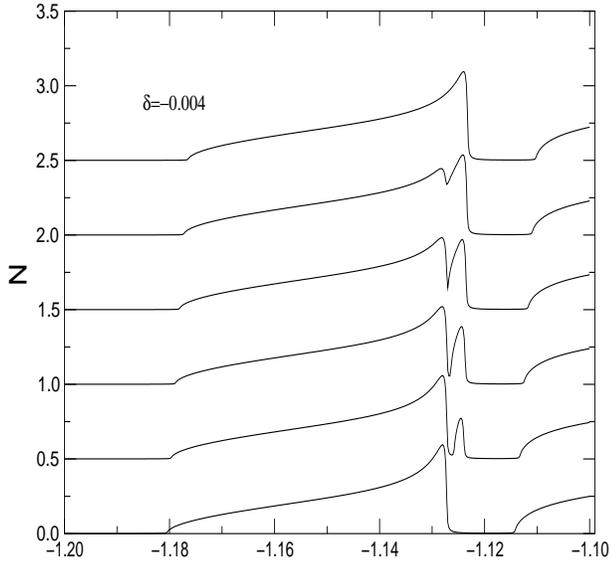}
\caption[fig7]
{Density of states $N(\omega)$ for the Holstein alloy 
in the vicinity of the first polaronic subband. The polaron 
parameters are $\lambda=1.96$ and $\alpha=2.8$ [see 
Table \ref{table}] and 
$\eta=10^{-4}$.
The scattering strength $\delta=-0.004$ 
and the concentration of the B atoms varies from 
$c=0.0$ (bottom) to $c=1.0$ (top) in steps of $\Delta c=0.2$. 
For clarity the density of states corresponding to different
concentrations are artificially shifted along the vertical axis.
}
\label{fig7}     
\end{figure}

\begin{figure}[t]
\psfig{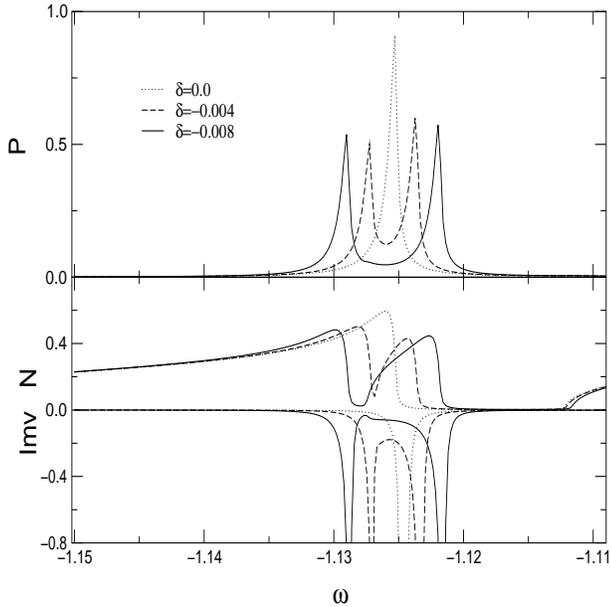}
\caption[fig8]
{The lower panel shows the density of states $N(\omega)$ and
the imaginary part of the coherent potential $Imv(\omega)$
for the Holstein alloy in the vicinity of the high-energy edge of
the first polaronic
subband for $c=0.5$ and three different values of the
scattering strength $\delta$.
The polaron
parameters are $\lambda=1.96$ and $\alpha=2.8$ [see
Table \ref{table}] and
$\eta=10^{-4}$. The corresponding
spectrally resolved return probabilities $P(\omega,\eta)$
are given in the upper panel.   
}  
\label{fig8}
\end{figure}

\begin{figure}[t]
\psfig{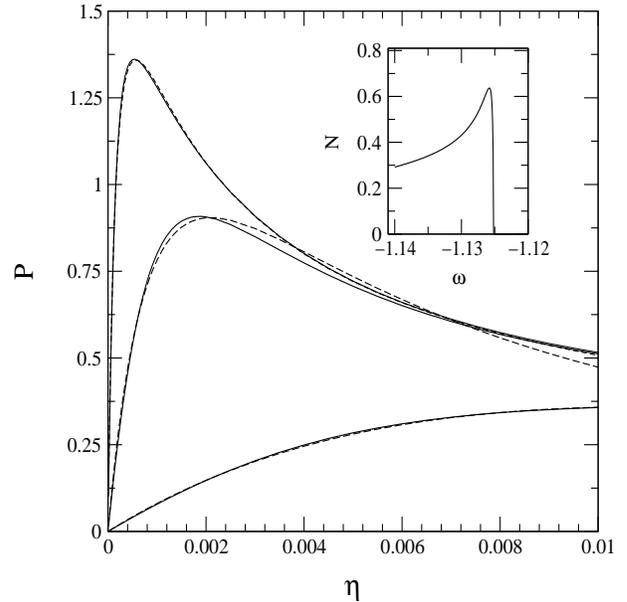}
\caption[fig9]
{$\eta-$asymptotics of the spectrally resolved return probability 
$P(\omega,\eta)$ for the ordered Holstein
model. The polaron parameters are $\lambda=1.96$ and $\alpha=2.8$
[see Table \ref{table}]. The three sets of curves correspond 
to three energies: $\omega=-1.135, -1.127$, and  
$-1.12549$ (from bottom to top).
Solid [dashes] lines depict the
numerical data [asymptotic expansion given in
Eq. (\ref{asympexp})]. The corresponding density of states for 
$\eta=10^{-8}$ is shown in the inset.
}
\label{fig9}
\end{figure}

\begin{figure}[t]
\psfig{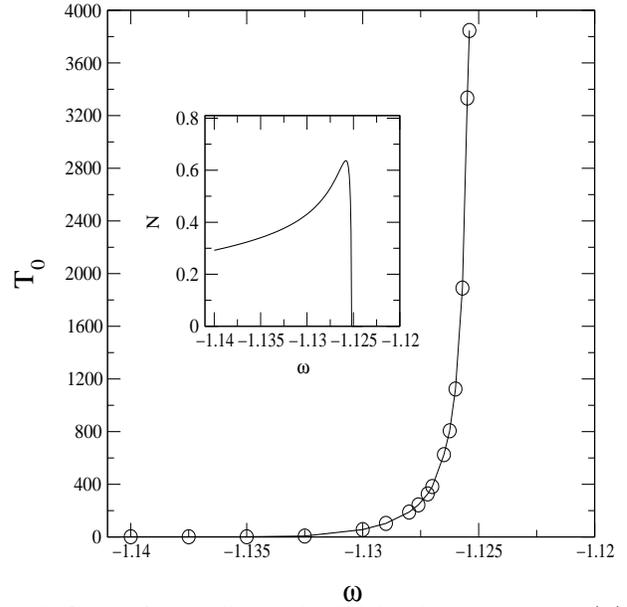}
\caption[fig10]
{Spectrally resolved delocalization time $T_0(\omega)$ for the ordered Holstein
model in the vicinity of the 
high energy edge of the first polaronic subband. The polaron  
parameters are $\lambda=1.96$ and $\alpha=2.8$ 
[see Table \ref{table}]. The inset shows the density of states 
for $\eta=10^{-8}$. Note, the data for $T_0(\omega)$ are for
energies below the $\eta\rightarrow 0$ high-energy edge.}
\label{fig10}
\end{figure}

\begin{figure}[t]
\psfig{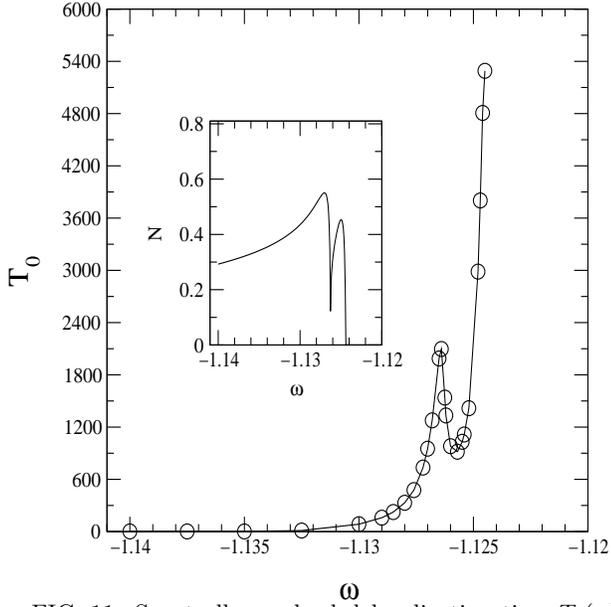}
\caption[fig11]
{Spectrally resolved delocalization time $T_0(\omega)$ for the Holstein alloy
in the vicinity of the
high energy edge of the first polaronic subband for $c=0.5$ 
and $\delta=-0.002$. The polaron
parameters are $\lambda=1.96$ and $\alpha=2.8$
[see Table \ref{table}]. The inset shows the density of states 
for $\eta=10^{-8}$. Note, the data for $T_0(\omega)$ are for
energies below the $\eta\rightarrow 0$ high-energy edge.}   
\label{fig11}
\end{figure}

The physical content of $P(\omega,\eta)$ can be extracted 
from Figs. \ref{fig4}c, \ref{fig5}d, \ref{fig6}c, and
the upper panel of Fig. \ref{fig8}, where we plot, for the 
model parameters given in the respective captions,  
$P(\omega,\eta)$ for $\eta=10^{-4}$. All the 
data show that $P(\omega,\eta)$ does not uniformly (in $\omega$)
approach zero as $\eta\rightarrow 0$. Instead, $P(\omega,\eta)$
is substantially enhanced for certain energies $\omega$. 
Close inspection of the data shows, moreover, that 
the maximum of $P(\omega,\eta)$ always occurs 
slightly below the high-energy edges of the 
respective (mini-)subbands, i.e., in the spectral 
regions where the dispersions become extremely flat.
(This can be most clearly seen in Fig. \ref{fig5}.)  
The enhancement of $P(\omega,\eta)$ is caused by the band flattening
and is therefore a direct measure of the sluggishness of 
the corresponding states due to their large phonon admixture. 
More precisely, because $1/\eta$ can be interpreted as a 
characteristic time scale, a 
large value of $P(\omega,\eta)$ implies 
that the electron in the state at energy $\omega$ is not yet
delocalized on the time scale $1/\eta$. 
Thus, the $\eta-$asymptotics of 
$P(\omega,\eta)$, i.e., the way $P(\omega,\eta)$ approaches 
zero for $\eta\rightarrow 0$, reveals the time scale on which 
the state at energy $\omega$ is {\it temporarily} trapped.
[Note, once more, we study
the $\eta-$asymptotics of $P(\omega,\eta)$ only for $\omega$ 
below the $\eta\rightarrow 0$ high-energy edge and not for 
energies belonging to the artificial tail which is present 
for a finite $\eta$.]

In what follows we put these qualitative
considerations on a mathematical basis.    
To that end, we first note that Eq. (\ref{returnprob1}) suggests to 
interpret $f(\omega-i\eta,\omega+i\eta)$ as a ``spectral density''
corresponding to $p(2\eta)$, the Laplace image of $1/2P(t/2)$.
That is, $f(\omega-i\eta,\omega+i\eta)$ gives the {\it spectrally 
resolved} diffusion behavior of the electron in the $\eta$ domain.
Accordingly, the counterpart of $f(\omega-i\eta,\omega+i\eta)$ in 
the time domain, which we denote by $F(\omega,t)$, 
is the ``spectral density'' corresponding 
to $1/2P(t/2)$. That is, it gives the
{\it spectrally resolved} diffusion behavior of the electron in the 
time domain. 

That the $\eta$-asymptotics of $P(\omega,\eta)$ indeed contain  
the spectrally resolved diffusion behavior of the electron
can be clearly seen in Fig. \ref{fig9}, 
where, for the ordered Holstein model with the polaron parameters given
in Table \ref{table}, $P(\omega,\eta)$ is plotted as a function of 
$\eta$ for three representative 
energies $\omega$ within the lowest polaronic subband, i.e., 
for energies below the $\eta\rightarrow 0$ high-energy edge. 
For the Holstein alloy, the overall
behavior of $P(\omega,\eta)$ is the same. All three curves feature 
a hump whose shape and position strongly depend on $\omega$. For  
$\omega$ in the immediate vicinity to the high-energy edge of the polaronic 
subband, the hump appears at a very small
$\eta$ and is very sharp. Far inside the subband, in contrast, the 
maximum appears at a larger $\eta$ and is less pronounced.  
The distinct $\omega$ dependence of the hump indicates 
different charateristic time scales for the
diffusion behavior of the respective states. 

To explicitly extract this time scale, we perform an asymptotic expansion
of $P(\omega,\eta)$ for small $\eta$. The data shown in Fig. \ref{fig9}
suggest making the following ansatz 
\begin{eqnarray}
P(\omega,\eta) \approx K(\omega)
{{\eta\over\eta_0(\omega)}\over{(1+{\eta\over\eta_0(\omega)})^{1+\nu(\omega)}}},  
\label{asympexp}
\end{eqnarray}
and determining the parameters $K(\omega)$,  
$\eta_0(\omega)$, and $\nu(\omega)$
through a least-square fit. The dashed lines in Fig. \ref{fig9}
show that this approach works reasonably well. 
Using the definition of $P(\omega,\eta)$
given in Eq. (\ref{returnprob2}) and employing a
general mathematical theorem about Laplace transforms
\cite{Doetsch89}, we obtain  
the asymptotic behavior of $F(\omega,t)$ for
large $t$ [note, since we measure energy in units of 
$2J$, time is mearured in units of $(2J)^{-1}]$: 
\begin{eqnarray}
F(\omega,t) \approx
K(\omega){{(t\eta_0(\omega))^{\nu(\omega)}}\over{2\Gamma(1+\nu(\omega))}}e^{-t\eta_0(\omega)},
\end{eqnarray} 
where $\Gamma(x)$ denotes the Gamma function. 

From this equation we infer that the parameter $\eta_0(\omega)$
gives rise to a 
characteristic time scale. Defining 
$T_0(\omega)=1/\eta_0(\omega)$, we see, in particular,
that for $t \gg T_0(\omega)$ the function $F(\omega,t)$ decays 
exponentially to zero.
For $t \ll T_0(\omega)$, however, $F(\omega,t)$ is 
finite, implying that the state at energy $\omega$ is not yet 
delocalized. It is therefore natural to interprete $T_0(\omega)$ as a 
(spectrally resolved) {\it delocalization 
time} characterizing {\it temporary} trapping, i.e., $T_0(\omega)$ 
quantifies the ``sluggishness'' of the state at energy $\omega$. 

In Fig. \ref{fig10} we plot $T_0(\omega)$ 
for the ordered Holstein model 
in the vicinity of the high-energy edge
of the lowest polaronic subband.  
The polaron parameters are the same as in Fig. \ref{fig4} 
(see Table \ref{table}). In the immediate vicinity of the
high-energy edge, where the dispersion becomes extremely
flat due to the high phonon admixture in the states, the  
delocalization times increase roughly three orders of
magnitude (but do not diverge). That is, 
the states in the flat part of the polaronic subband
appear to be temporarily trapped on the time scales where 
the rest of the 
states are already delocalized. 
Clearly, the band flattening and the
enhanced delocalization times are correlated and 
have, moreover, the same microscopic origin, namely the 
increased phonon admixture in the states at the high-energy
edge of the subband. More formally, it is the 
{\it divergence-induced} strong $\omega-$dependence 
of $Rev(\omega)$ for  
$\omega_m\le\omega\le\omega_h$ which causes 
the large delocalization times at the high-energy edge
of the lowest polaronic subband. 

At this point it is appropriate to comment on the work of 
Hotta and Takada \cite{HT96} who suggested that for energies where
the self-energy (or, in our case, the coherent potential)
diverges, states would be trapped 
because of the infinitely strong electron-phonon coupling. 
They studied the half-filled (ordered) Holstein-Hubbard model, but
their conclusion is quite general and should, if valid, hold
for any model which produces divergences in the self-energy.
Such divergences are, however, always inside a gap. In fact,
they signal the presence of a gap, as can be seen, e.g., in 
Figs. \ref{fig4} and \ref{fig5}, that is, 
there are no states 
at the divergence energy which could be trapped. 
In contrast, our investigation suggests that, it is the effect the divergence 
has on states
{\it within} the spectrum, namely, the 
band flattening it causes below
the high-energy edge of the lowest polaronic subband, which could 
perhaps induce trapping in that particular energy range.
Within the DCPA, however,
the effect is too weak, giving rise only to large but finite 
delocalization times and, hence, only to temporary trapping. 

It has been also suggested \cite{C97} that, if, due to some mechanism, 
a finite density of states appears at the divergence energy, 
the corresponding states would be trapped. A finite $\eta$ 
simulates such a mechanism. 
However, as can be seen in Fig. \ref{fig5}, 
the maximum of $P(\omega,\eta)$
does not occur in the tail 
but roughly at the maximum of the subband density of states.
Therefore, at least within the DCPA, 
it is the band flattening and {\it not} the tail 
which is responsible for large delocalization times.

The concept of the delocalization time provides an
appealing explanation of the spectral changes induced by 
alloying. In the previous subsection we have seen that the 
high-energy edge of the polaronic subband responds much more
strongly to alloying than the bottom of the subband. From
the perspective of the delocalization time this is now clear, 
because the high-energy edge contains ``sluggish''
defect-like states, whose 
large delocalization times make them feel the 
randomness of the on-site potential particularly strongly.  

The delocalization times are also modified due to alloying. This is
shown in Fig. \ref{fig11} for a particular set of alloy parameters,
$c=0.5$ and $\delta=-0.002$. The polaron parameters
are the same as before (see Table \ref{table}). In the 
previous subsection we have seen that the main effect of alloying is 
the formation of two mini-subbands with, due to 
the large phonon admixture in the states, rather flat 
dispersions at the two high-energy edges. 
As a consequence, 
the delocalization times are now enhanced in two narrow energy 
regions: At the high-energy edge of the A mini-subband and at the
high-energy edge of the B mini-subband. 
Note, as in the 
case without disorder, the enhancement of the delocalization
times
is due to the band flattening effect, i.e., it
is driven by strong on-site electron-phonon correlations. Even
with the assistance of electron-impurity scattering,   
the on-site correlations are however not strong enough to 
produce divergent delocalization times. 

The numerical results for the delocalization times suggest
that we may tentatively distinguish two types
of states: ``Fast'' {\it quasi-partice-like} states
and ``sluggish''
{\it defect-like} states. In the 
ordered Holstein model, defect-like
states appear at the high-energy edge
of the polaronic subband. Whereas in the Holstein alloy, 
defect-like states occur at the high-energy edges of the 
two mini-subbands. With and without disorder,
strong on-site 
electron-phonon correlations are responsible for the 
formation of defect-like states. Because
the delocalization times of the defect-like states are 
{\it several orders} of magnitude larger than the delocalization
times of the quasi-particle-like states, it is moreover 
conceivable that the sluggishness of the defect-like 
states may anticipate the onset of trapping. 

\section{Conclusions}

We have provided a self-contained description of the DCPA and 
employed it to investigate the dynamics  
of a single electron in the Holstein model augmented by  
site-diagonal, binary-alloy type disorder (Holstein alloy).
Using multiple-scattering theory, the conventional CPA technique, 
we derived the DCPA equations for the averaged two- and four-point
functions and emphasized the main approximation involved: namely,
after each hop from one lattice site to another the electron's
spatial memory is erased. We visualized the lack of spatial 
memory in terms of self-avoiding paths and employed this picture
to give an intuitive explanation why the DCPA becomes exact for
a system with infinite coordination number. 

Our numerical results focused on the intermediate 
electron-phonon coupling
regime, where polaronic subbands start to emerge. We 
investigated, for representative parameter sets, the effect 
of alloying on the high-energy edge of the lowest 
polaronic subband. These states are very susceptible to 
impurity scattering because of their large phonon admixture, 
which leads to a flat dispersion in this part of the spectrum, 
and, as a consequence, to a steeple-like density of states. 
The observed modifications of the spectral properties 
are therefore reminiscent of typical CPA results 
obtained for a model with a steeple-like density
of states: A lack of $c \leftrightarrow 1-c$ symmetry, 
an appearance of mini-subbands in the vicinity of the steeple
(i.e., high-energy edge) of the subband for scattering strengths 
much smaller
than the subband width, and an asymmetry in the 
component density of states. 

The most notable 
result is, however, the large enhancement of the 
spectrally resolved return probability $P(\omega,\eta)$ 
for small but finite $\eta$ in the narrow energy regions 
where the dispersion of the polaron states becomes extremely
flat. This occurs in the absence (presence) of alloy-type
disorder at the high-energy edge(s) of the 
polaronic subband (mini-subbands). To elucidate the 
physical meaning of this strong enhancement, we 
analyzed the $\eta-$asymptotics of $P(\omega,\eta)$ and 
introduced the concept of a 
spectrally resolved delocalization time which 
is the characteristic time scale on which, at a given
energy, the electron leaves a given site. The strong 
enhancement of $P(\omega,\eta)$ in the flat part(s) of 
the subband (mini-subbands) signals large delocalization 
times.
According to their delocalization times we could thus
classify, within the limitations of the DCPA, ``fast''
quasi-particle-like polaron states at the bottom 
and ``sluggish'' defect-like polaron
states at the top of the subband (mini-subbands). 

Within the DCPA, we could not decide, however, whether
``defect-like'', i.e., temporarily trapped
states are in fact ``defect'', i.e., trapped states.
At this point it is important to recall that
within the DCPA only on-site processes contribute to the
delocalization times. In contrast, inter-site processes
are completely neglected. Particularly, back-scattering
processes are expected to 
have a significant effect.
In the conventional alloy problem, for example, these
processes even yield, depending on the model parameters,
infinite delocalization times at certain energies. It is therefore
conceivable that the DCPA results for the delocalization
times of polaron states presented in this paper
are only lower bounds 
and temporarily trapped states may in fact
be trapped states. More sophisticated methods
beyond the DCPA, taking back-scattering and 
other non-mean field effects explicitly into
account, are clearly needed to further elucidate the
trapping issue.           

\section*{Acknowledgments}

One of the authors (F.X.B.) would like to thank Prof. Dr. H. B\"ottger for 
several valuable
discussions during the course of this work. Useful and stimulating 
conversations with 
Dr. Holger Fehske are also very much appreciated. F.X.B. also acknowledges 
the hospitality of
the Theoretical Division at Los Alamos National Laboratory where 
this work was initiated. This work
is supported in part by the U.S. Department of Energy.

\appendix

\section{Polaron Impurity Model (PIM)}

In this appendix we show how 
the local two-point function
${\cal G}_{ii}(z)$ can be calculated from the PIM. 
Towards that end, we apply $<i|(N,n_i|...|m_i,N)|i>$ on both sides 
of Eq. (\ref{pim2}) and obtain in compact matrix notation
\begin{eqnarray}
{\cal D}^{(N)}(z)&=&\tilde{G}^{(N)}(z) + \tilde{G}^{(N)}(z)
\nonumber\\
&\times& \left [\sigma_{imp}^{(N)}(x_i;z) +
\sigma_{ph}^{(N)}(g) \right]
\nonumber\\
&\times& {\cal D}^{(N)}(z),
\label{Dmatrix}
\end{eqnarray}           
with matrices
\begin{eqnarray}
{\cal D}^{(N)}_{nm}(z)&=&<i|(0,n_i|D^{(N)}_i(z)|m_i,0)|i>,
\label{Dmatrixnm}\\
\left[\sigma^{(N)}_{imp}(x_i;z)\right]_{nm}&=&
\left[x_i\delta-v^{(N+n)}(z)\right]\delta_{n,m},
\\
\left[\sigma^{(N)}_{ph}(g)\right]_{nm}&=&
-g\sqrt{n+1}\delta_{m,n+1}
-g\sqrt{n}\delta_{m,n-1},
\\
\left[\tilde{G}^{(N)}(z)\right]_{nm}&=&
{\cal G}_{ii}^{(N+n)}(z)\delta_{n,m}.
\end{eqnarray}
To derive Eq. (\ref{Dmatrix}) we employed 
$\Delta H_i(z)\sim |i><i|$
together with
${\cal G}_{ii}^{(q)}(z)=<i|g^{(q)}(z)|i>$,
which follows from the definition of $G_{eff}(z)$. 

If we now introduce an auxiliary matrix 
\begin{eqnarray}
\tilde{H}^{(N)}(z)=\left[1 - \tilde{G}^{(N)}(z)\sigma_{ph}^{(N)}(g) \right]^{-1} \tilde{G}^{(N)}(z),
\end{eqnarray}
which sums all electron-phonon scattering processes, Eq. (\ref{Dmatrix}) can be rearranged into
\begin{eqnarray}
{\cal D}^{(N)}(z)=
\left[1 - \tilde{H}^{(N)}(z) \sigma_{imp}^{(N)}(x_i;z) \right]^{-1} \tilde{H}^{(N)}(z).
\label{auxA}
\end{eqnarray}
The configuration average is straightforwardly performed since (single-site)
randomness enters only through $\sigma_{imp}^{(N)}(x_i;z)$. We find,
using the bi-modal probability distribution $p(x_i)=c\delta(x_i)+(1-c)\delta(x_i-1)$, 
\begin{eqnarray}
\langle\langle {\cal D}^{(N)}(z) \rangle\rangle
&=&\int dx_i p(x_i) {\cal D}^{(N)}(z)
\nonumber\\
&=& (1-c) \left[1 - \tilde{H}^{(N)}(z)\sigma_{imp}^{(N)}(1;z) \right]^{-1}
\nonumber\\
&\times& \tilde{H}^{(N)}(z)
\nonumber\\
&+&c \left[1 - \tilde{H}^{(N)}(z)\sigma_{imp}^{(N)}(0;z) \right]^{-1} 
\nonumber\\
&\times&\tilde{H}^{(N)}(z)
\nonumber\\
&=&(1-c){\cal D}^{A,(N)}(z) + c {\cal D}^{B,(N)}(z),
\end{eqnarray}
or, explicitly in terms of matrix elements,
\begin{eqnarray}
\langle\langle{\cal D}^{(N)}_{nm}(z)\rangle\rangle=(1-c){\cal D}_{nm}^{A,(N)}(z) +
c {\cal D}_{nm}^{B,(N)}(z),
\end{eqnarray}
where the amplitudes ${\cal D}_{nm}^{\lambda,(N)}(z)$ satisfy a set of 
recursion relations
($\lambda=A,B$)
\begin{eqnarray}
{\cal D}_{nm}^{\lambda,(N)}(z)&=&F^{\lambda,(N+n)}(z)\delta_{n,m}
-gF^{\lambda,(N+n)}(z)
\nonumber\\
&\times& \left[\sqrt{n+1}{\cal D}_{n+1m}^{\lambda,(N)}(z)
+\sqrt{n}{\cal D}_{n-1m}^{\lambda,(N)}(z)\right],
\label{RR}
\end{eqnarray}
with 
\begin{eqnarray}
&F&^{\lambda,(n)}(z)=
\nonumber\\
& &{1 \over {z-n\Omega-\epsilon_\lambda
+{\cal G}_{ii}^{-1}(z-n\Omega)
-{\cal R}[{\cal G}_{ii}(z-n\Omega)]}}.
\end{eqnarray}

From the self-consistency condition, Eq. (\ref{sce6}), finally follows 
\begin{eqnarray}
{\cal G}_{ii}(z)&=&\langle\langle {\cal D}_{00}^{(0)}(z)\rangle\rangle
\nonumber\\
&=&(1-c){\cal D}_{00}^{A,(0)}(z) + c {\cal D}_{00}^{B,(0)}(z),
\end{eqnarray}
which directly yields Eq. (\ref{cfraction}) 
if the 
continued fraction expansion \cite{C84} for the 
amplitudes ${\cal D}_{00}^{\lambda,(0)}(z)$ is invoked to 
iteratively solve  
Eq. (\ref{RR}).  

\section{Derivation of Eq. (\ref{F3})}

In this appendix, we bring Eq. (\ref{f2})
into the numerically more convenient 
form of Eq. (\ref{F3}).
As a preparatory step we 
notice, using the definition of the PIM, Eq. (\ref{pim1}), 
that the matrix elements of the atomic T-matrix defined in Eq. 
(\ref{tnm}) can 
be written as
\begin{eqnarray}
t^{(0)}_{0q}(z)={{{\cal D}^{(0)}_{0q}(z)-\delta_{0q}{\cal G}_{ii}(z)}
\over{{\cal G}_{ii}(z){\cal G}_{ii}(z-q\Omega)}},
\label{auxD1}
\end{eqnarray}
with ${\cal D}^{(0)}_{0q}(z)$ defined in Eq. 
(\ref{Dmatrixnm}); here, we employed again the identity  
${\cal G}_{ii}^{(q)}(z)=<i|g^{(q)}(z)|i>$.
If we now insert Eq. (\ref{auxD1}) into Eq. ($\ref{f2}$) and take into account that
$\langle\langle{\cal D}_{00}^{(0)}(z)\rangle\rangle={\cal G}_{ii}(z)$,
Eq. (\ref{f2}) reduces to 
\begin{eqnarray}
f(z_1,z_2)=\sum_{q=0}^\infty
\langle\langle {\cal D}_{0q}^{(0)}(z_1){\cal D}_{q0}^{(0)}(z_2)\rangle\rangle.
\end{eqnarray}
For $z_1=z_2^*=z^*$, we may use 
$v^{(0)}(z)=[v^{(0)}(z^*)]^*$ 
and slightly rearrange the above equation into
\begin{eqnarray}
f(z^*,z)=\sum_{q=0}^\infty
\langle\langle |{\cal D}_{q0}^{(0)}(z)|^2 \rangle\rangle.
\label{auxD2}
\end{eqnarray}
Employing now the matrix notation introduced in appendix A, 
the configuration average of
Eq. (\ref{auxD2}) can be performed explicitly. 
To be specific, from Eq. (\ref{auxA}) it follows that   
\begin{eqnarray}
{\cal D}_{q0}^{(N)}(z)&=&
\sum_r \{ \left[1 - \tilde{H}^{(N)}(z)\sigma_{imp}^{(N)}(x_i;z) \right]^{-1} \}_{qr}
\nonumber\\
&\times& \tilde{H}_{r0}^{(N)}(z),
\end{eqnarray}
which, using the bi-modal probability distribution   
$p(x_i)=c\delta(x_i)+(1-c)\delta(x_i-1)$, immediately yields
\begin{eqnarray}
\langle\langle |{\cal D}_{q0}^{(0)}(z)|^2 \rangle\rangle
&=&\int dx_i~ p(x_i) |{\cal D}_{q0}^{(0)}(z)|^2
\nonumber\\
&=& (1-c) |{\cal D}_{q0}^{A,(0)}(z)|^2
\nonumber\\
&+& c |{\cal D}_{q0}^{B,(0)}(z)|^2,
\label{auxD3}
\end{eqnarray}
with the amplitudes ${\cal D}_{q0}^{\lambda,(0)}(z)$ satisfying the recursion relations
Eq. (\ref{RR}).
Combining Eqs. (\ref{auxD2}) and (\ref{auxD3}) and setting $z=\omega+i\eta$, we 
finally obtain
Eq. (\ref{F3}).



\newpage

\begin{table}
\caption[table1]
{Polaron model parameters (in units of $2J$)}
\begin{tabular}{ccccc} 
J & $\Omega$ & g & $\lambda=2g^2/\Omega$ & $\alpha=g/\Omega$ \\ \hline
0.5 & 0.125 & 0.35 & 1.96 & 2.8 
\label{table}
\end{tabular}
\end{table}

\end{document}